\documentclass{phb-proc4-auth}

\usepackage{graphicx}
\usepackage{amssymb}

\begin{document}
\begin{frontmatter}


\journal{SCES '04}


\title{Doping dependence of density response and
bond-stretching \\ phonons in cuprates }

%
%
%
%

%
%

\author{Peter Horsch\corauthref{1}}
\author{Giniyat Khaliullin}

%

\address{Max-Planck-Institut f\"ur Festk\"orperforschung, 
D-70569 Stuttgart, Germany}

%
%
%
%


%
%
%
%

\corauth[1]{Corresponding Author: Max-Planck-Institut FKF, 
D-70569 Stuttgart, Germany. Phone: +49 711 689-1550 
Fax: +48 711 689-1702, Email: P.Horsch@fkf.mpg.de }


\begin{abstract}

We explain the anomalous doping dependence of zone boundary
$(\pi,0)$ and $(\pi,\pi)$ bond-stretching phonons in 
La$_{2-\delta}$Sr$_{\delta}$CuO$_4$ in the range $0<\delta<0.35$.
Our calculations are based on a theory for the density response
of doped Mott-Hubbard insulators.

\end{abstract}

%
%

\begin{keyword}

High temperature superconductors, LaSrCuO, 
Density response of doped Mott-Hubbard insulator,
bond-stretching phonons

\end{keyword}


\end{frontmatter}

%
%
%
%
%
The relevance of electron-phonon interaction for high-T$_c$ superconductivity
has been controversially discussed over the years. 
Certainly a clear signature of the
interplay of charge carriers and phonons is the strong dependence of
breathing and bond-stretching phonon modes as function of doping
as evidenced in neutron scattering data. 
This was first oberved in LaSrCuO and YBCO 
\cite{Pintschovius94,Braden99}, yet appears meanwhile as
a generic feature of all high-T$_c$ superconductors.

The theoretical study of phonon renormalization in strongly correlated
high-T$_c$ superconductors requires a calculation of the density response
for a doped Mott-Hubbard insulator (MHI).
Such a theory has been developed for the $t$-$J$ model based on a description of
correlated electrons in the framework of the slave boson approach \cite{Khaliu96}, 
as well as within the complementary slave fermion method \cite{Horsch00}, 
with results in favorable agreement with exact diagonalization 
studies \cite{Tohyama95}.

In this contribution we report the doping dependence for
La$_{2-\delta}$Sr$_{\delta}$CuO$_4$
in the range $0<\delta<0.35$ of the zone-boundary
$(\pi,\pi)$ breathing phonon and of the highly anomalous $(\pi,0)$ bond-stretching
mode, i.e., calculated by means of the slave boson approach \cite{Khaliu96}. 
The phonon data presented include theoretical results for 
optimal doping, $\delta=0.15$, 
published earlier \cite{Khaliu97}.

 \begin{figure}[h]
     \centering{
     \includegraphics[angle=0,width=4.0cm]{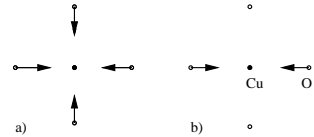}
}
     \caption{Displacements of O ions (a) for ${\bf q}=(\pi,\pi)$ and
(b) for the $(\pi,0)$ half-breathing mode.  } 
 \end{figure}
The phonon modes under discussion couple directly to the density of holes
$n_i^h=h_i^+h_i$ since the associated O-ion displacements lead to a modulation
of the Zhang-Rice singlet energy. This is also the reason why the renormalization
of these phonons can be treated in the framework of the $t$-$J$ model.
The change of the Zhang-Rice energy  $E_{ZR}=8 t_{pd}^2/\Delta\epsilon$
with respect to the oxygen displacements 
$u_{\alpha}^i$, $\alpha=x,y$, of the four O-neighbors at $R_i+\delta^O_{\alpha}$
around the Cu-hole at $R_i$ yields
the linear electron-phonon coupling \cite{Khaliu97}
\begin{equation} 
H_{e-ph}=
g\sum_i(u^i_{x}-u^i_{-x}+u^i_{y}-u^i_{-y}) h^+_ih_i.
\end{equation}
We assume that the resonance integral obeys the Harrison relation 
$t_{pd}\propto  r_0^{-7/2}$, where $r_0$ is the Cu-O distance, and obtain
$g=7 E_{ZR}/ 4 r_0$, i.e., $g\approx 2$eV/\AA. 
The lattice part of the Hamiltonian is determined
by the force constant $K\approx 25$eV/\AA$^2$ for the longitudinal O-motion.
Due to the structure of $ H_{e-ph}$ the bond stretching modes couple directly
to the density response function $\chi({\bf q},\omega)$. 
We have studied the renormalization of the
phonon Green's function 
\begin{equation}
D^{ph}_{{\bf q},\omega}=\frac{\omega_{{\bf q},0}}
{\omega^2-\omega^2_{{\bf q},0}(1-\alpha_{\bf q}
\chi({\bf q},\omega))}, 
\end{equation}
where $\omega_{{\bf q},0}$ is the bare phonon frequency, i.e. measured in the undoped
parent compound ( $\omega_{{\bf q},0}\sim80$ and 90 meV for $(\pi,0)$ and
$(\pi,\pi)$, respectively \cite{Pintschovius94}). 
The coupling function $\alpha_{\bf q}=\frac{4g^2}{K}(\sin^2{q_x/2}+\sin^2{q_y/2})$ 
 vanishes at the $\Gamma$ point and becomes maximal at the zone edges. 
Based on the parameters of the $pd$-model we estimate
for the dimensionless coupling constant $\xi=g^2/ztK \sim 0.07 - 0.12$.

 \begin{figure}
     \centering{
     \includegraphics[width=3.7cm]{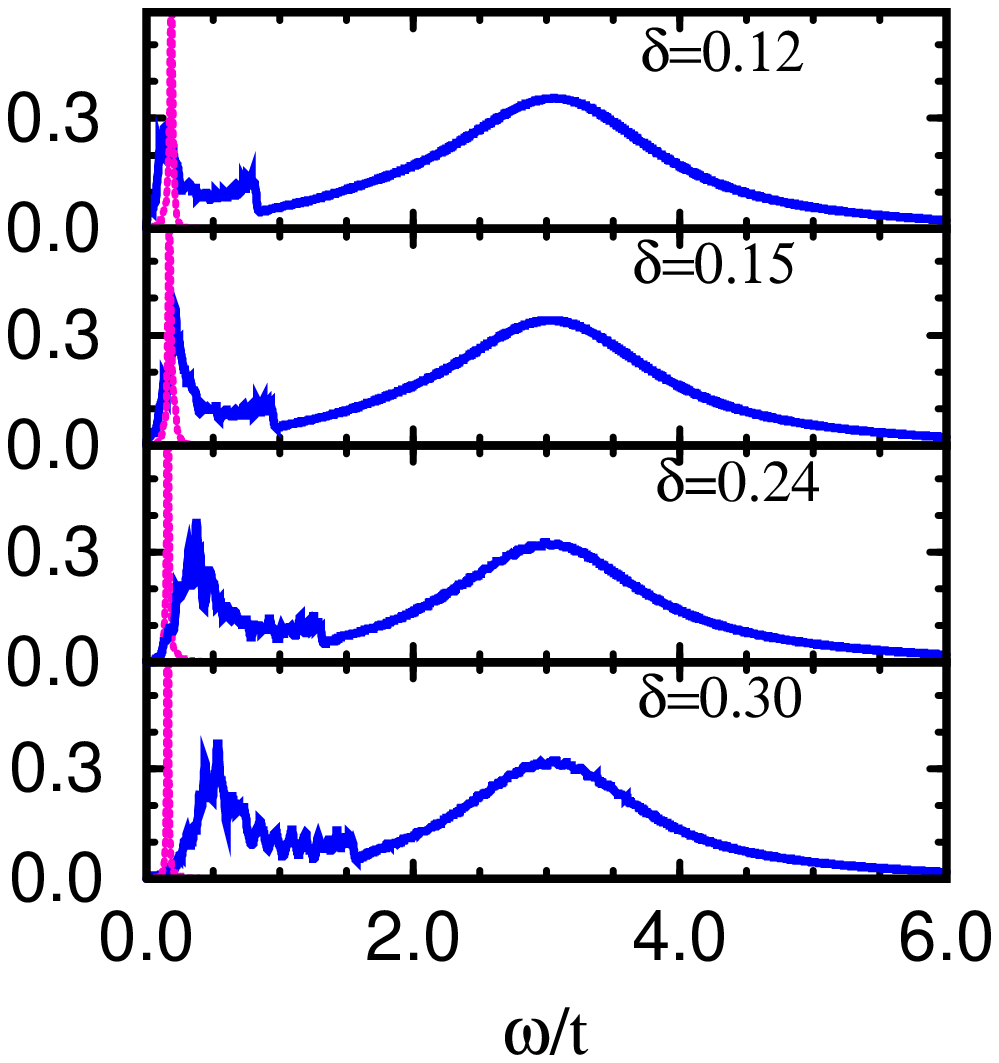}
     \includegraphics[width=3.7cm]{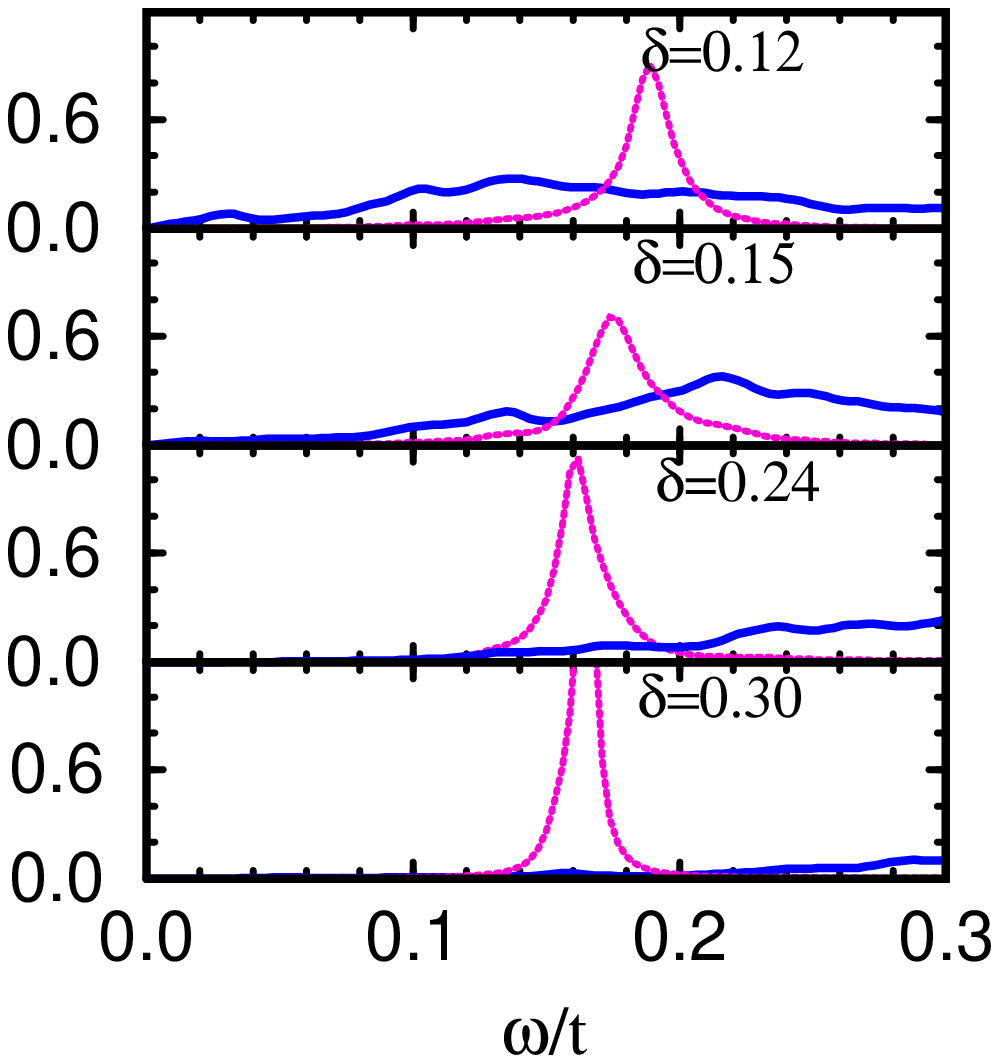}
}
     \caption{Dynamic charge structure factor $N({\bf q},\omega)$
({\it normalized by the density of holes $\delta$}) at momentum 
${\bf q}=(\pi,0)$ for different doping concentrations $\delta$.
(a) On the large energy scale revealing the scaling of the edge of the
spinon particle hole continuum $\sim 4(\kappa J +\delta t)$
and of the polaron peak $\sim (\kappa J +\delta t)$.
(b)  $N({\bf q},\omega)$ at low energy and 
doping dependence of the spectral function of the
${\bf q}=(\pi,0)$ bond stretching phonon.  } 
 \end{figure}

The central quantity controlling the phonon propagator, Eq.(2), is of course the
density response function 
$\chi({\bf q},\omega)=\langle \delta n^h \delta n^h\rangle_{{\bf q}\omega}$
of the doped Mott-Hubbard insulator, whose momentum
and frequency dependence was studied in Ref.\cite{Khaliu96}.
Figure 2(a) shows the doping dependence of the associated density fluctuation spectrum
$N({\bf q},\omega)=\frac{1}{\pi} \chi''({\bf q},\omega)/\delta$ for ${\bf q}=(\pi,0)$.
At ${\bf q}=(\pi,0)$ the main structure of $N({\bf q},\omega)$ lies at high energy
$\sim 3 t$
and is strongly broadened due to the scattering of holes by spin excitations.
The underlying polaronic mechanism leads at the same time 
to a peak in $N({\bf q},\omega)$ at low energy 
$\sim (\kappa J +\delta t)$ with $\kappa\sim 0.3$, 
which can be attributed to the coherent motion of the polarons, i.e.,
holes dressed by spin excitations moving coherently on the energy scale 
dictated by the spin degrees of freedom. 
This polaron peak, whose energy scales with $\delta$,
is the origin of the anomalous behavior of the $(\pi,0)$ bond-stretching phonon.
Interestingly this polaron structure is absent 
near $(\pi,\pi) $\cite{Tohyama95,Khaliu96}, hence no anomalous behavior
is expected for the breathing phonon. The origin of the anomalous behavior of the 
density response of the doped MHI near $(\pi,0)$ can be traced back to a strong 
$(\pi,0)\leftrightarrow (0,\pi)$ scattering mediated by spin-excitations with
momentum transfer close to  $(\pi,\pi)$. 
An additional feature seen in Fig. 2(a) is the spinon particle-hole continuum 
which at $(\pi,0)$ extends up to  $\sim 4(\kappa J +\delta t)$.

 \begin{figure}
     \centering
     \includegraphics[width=6.cm]{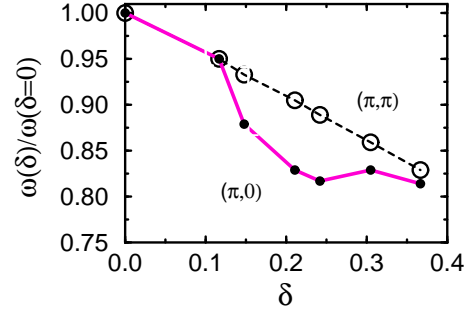}
     \caption{Predicted doping dependence  on the basis of the
$t$-$J$ model for $(\pi,\pi)$ breathing 
and $(\pi,0)$ bond-stretching
phonon energies in La$_{2-\delta}$Sr$_{\delta}$CuO$_4$.
Parameters as for Fig.2: 
$t=0.4$ eV, $J/t=0.3$, and dimensionless electron phonon
coupling constant $\xi=g^2/ztK \simeq 0.06$.} 
 \end{figure}

As $\chi({\bf q},\omega)$ is proportional to the hole density,  phonons are 
renormalized basically linearly with $\delta$, yet due to the polaron
structure at low energy along the $(\pi,0)$ direction, the $(\pi,0)$ phonon
reveals an anomalous energy shift and damping.
The strong doping dependence of this effect, shown in Fig.2(b) and Fig.3, is due to
the shift  $\propto(\kappa J + \delta t)$ of the polaron peak position 
in $N({\bf q},\omega)$ and its crossing of the bond-stretching phonon energy.   
Our results imply in particular that the large damping of the $(\pi,0)$
optical phonon near optimal doping  disappears at larger doping concentrations,
where the two phonon modes $(\pi,0)$ and $(\pi,\pi)$ behave similar.

In conclusion,  the anomalous renormalization and damping of bond stretching phonons
is a manifestation of the strongly correlated motion of holes in cuprates,
thus the study of these phonons provides a subtle test  
of the low-energy density response in cuprates at finite momentum transfer. 
%
%
%
%

%
%
  
%
%


\end{document}